\documentclass[11pt]{JHEP3}
\usepackage{epsfig}
\usepackage{amsfonts}
\usepackage{latexsym}
\usepackage{amsmath}
\usepackage{datetime}
\usepackage{mathrsfs}
\usepackage[square, comma, sort&compress,numbers]{natbib}

\numberwithin{equation}{section}

\newcommand{\bi}{\begin{itemize}}
\newcommand{\ei}{\end{itemize}}
\newcommand{\nn}{\nonumber}

\newcommand{\bea}{\begin{eqnarray}}
\newcommand{\eea}{\end{eqnarray}}
\newcommand{\be}{\begin{equation}}
\newcommand{\ee}{\end{equation}}

\title{Entanglement entropy of subtracted geometry \\ black holes}

\maketitle
\preprint{UPR-1261-T }
\author{\large Mirjam Cveti\v c$^{\dag,\ddag}$, Zain H. Saleem $^\dag$ and Alejandro Satz$^{\dag}$  

\email{cvetic@physics.upenn.edu},
\email{zains@sas.upenn.edu},
\email{alesatz@sas.upenn.edu}.

\vspace{0.5cm}

{\it $^\dag$ \normalsize{Department of Physics and Astronomy, University of Pennsylvania,}   \vspace{-2 mm} \\

\hspace{-0.3 cm} \normalsize{Philadelphia, PA 19104-6396, USA}
}

 \vspace{ 3 mm}

{\it $^\ddag$ \normalsize{Center for Applied Mathematics and Theoretical Physics},  \vspace{-2 mm}\\
 \hspace{-0.3 cm} \normalsize{University of Maribor, Maribor, Slovenia}
 }

  }

\abstract{

\vskip 5 mm

We compute the entanglement entropy of minimally coupled scalar fields on subtracted geometry black hole backgrounds, focusing on the logarithmic corrections. We notice that matching between the entanglement entropy of original black holes and their subtracted counterparts is only at the order of the area term. The logarithmic correction term is not only different but also, in general, changes sign in the subtracted case. We apply Harrison transformations to the original black holes and find out the choice of the Harrison parameters for which the logarithmic corrections vanish.
}

\begin{document}

\newpage
\section{Introduction}

Subtracted geometry black holes are obtained when one omits certain terms from the warp factor of the metric of general black holes \cite{CL11I,Cvetic:2011dn,Cvetic:2012tr}. The omission of these terms allows one to write the wave equation of the black hole in a completely seperable way \cite{Cvetic:1997xv},\cite{Cvetic:1997uw} and one can explicitly see that the wave equation of a massless scalar field in this slightly altered background of a general multi-charged rotating black hole acquires  an 
SL(2,R)$\times$SL(2,R)$\times$SO(3) 
symmetry. The reason why the "subtracted limit" is considered an appropriate limit for studying the internal structure of the
black hole is that the new subtracted geometry black holes have
the same horizon area and periodicity  of 
the angular and time coordinates in the near horizon
regions as the original black hole geometry it was constructed from.
The new geometry is asymptotically conical and is physically similar to that of a black hole in an asymptotically confining box \cite{Cvetic:2012tr}.\\

In this paper we will be studying the entanglement entropy of minimally coupled scalar fields across the horizon of these subtracted geometry black holes. Entanglement entropy, sometimes also referred to as  ``geometric entropy'', was introduced in high energy physics for the first time in \cite{Srednicki:1993im}, where  the reduced density matrx
and the corresponding entropy were directly calculated in the flat spacetime by tracing over the degrees of freedom
residing inside an imaginary surface. This entropy was shown to be propotional to the area of the entangling surface. This approach was later applied to black holes in \cite{Frolov:1993ym} to calculate the entanglement entropy of the quantum fields across the black hole horizon. In \cite{Susskind:1993ws}, an efficient computational method for calculating the entanglement entropy, called the ``replica trick'', was introduced. This method helped in developing a systematic method for calculating the UV divergent terms in the entanglement entropy. Here we will be following the same method to calculate the leading order plus the logarithmic corrections to geometric entropy of those black holes that arise as a result of the subtraction procedure applied to N=2 STU black holes \cite{cvetic1}. We will compare these results to that of the original black holes. Since the computation is performed across the black hole horizon, whose area is unaffected by the subtraction process, we expect to see a complete match at the tree level. However, we find that logarithmic corrections are different from the original black holes and therefore arrive to the conclusion that subtracted limit seems to be an appropriate limit only when the black hole is not too massive so that the logarithmic corrections are ignorable compared to the leading result.\\

We will also be studying the effect of Harrison tranformations on the entanglement entropy formulae we obtain. The Harrison transformations are certain infinite boost transformations that connect the four dimensional black holes to their subtracted counterparts while keeping the Cardy-like form of their entropy intact. In \cite{Cvetic:2012tr} the interpolating solution corresponding to the Schwarzchild black holes was constructed using these transformations and in \cite{Vir} this was done for the case of general four charge black holes. In one dimension higher, the subtracted geometry black holes lift to $AdS_3\times S^2$. In \cite{Jottar} it was suggested that the Harrison transformations are related to turning on some irrelevant operators in the dual CFT. In \cite{Cvetic:2013cja} these ideas were further elucidated and it was shown how this subtraction procedure using the Harrrison transformations along with some scaling transformations could be related to a simple combination of T-dualities and Melvin twists in the string theory framework. In this paper we will be mainly concerned with the behavior of the logarithmic corrections to the entanglement entropy of the Schwarzchild black hole under these transformations. One of the interesting universal features of the logarithmic corrections to the subtracted geometry black holes is that they have a different sign than the logarithmic corrections to the original black holes. Thus one is lead to believe that an interpolating black hole solution, using the Harrison transformations can be constructed, for which the logarithmic corrections vanish for certain choices of the Harrison parameters. We will study this interpolating solution for the easiest case of the Schwarzchild black hole. \\

The paper is organized as follows. In section 1 we introduce the subtracted black hole geometry. In section 2 we write explain how the calculations are performed and then write down the results for the original as well as subtracted black hole cases. In section 3 we explain the Harrison transformations and how they effect the entanglement entropy formulae for the case of static black holes. In the last section we summarize the main results and their significance.

\section{Original and subtracted black hole geometry}
The black holes we will be concerned with in this paper are solutions of the  bosonic sector four-dimensional Lagrangian density  of the
${\cal N}=2$ supergravity coupled to three vector supermultiplets. The original four-charge rotating solution was constructed in \cite{cvetic1} and the explicit expressions for all the four gauge fields was given in \cite{CCLPI}. The Lagrangian density of this theory is given by
\begin{align}
{\cal L}_4 = & R\, {*{\bf 1}} - \frac{1}{2} {*d\varphi_i}\wedge d\varphi_i 
   - \frac{1}{2} e^{2\varphi_i}\, {*d\chi_i}\wedge d\chi_i - \frac{1}{2} e^{-\varphi_1}\,
\big( e^{\varphi_2-\varphi_3}\, {*  F_{1}}\wedge   F_{ 1}\nn\\
 &+ e^{\varphi_2+\varphi_3}\, {*   F_{ 2}}\wedge   F_{ 2}
  + e^{-\varphi_2 + \varphi_3}\, {*  {\cal F}_1 }\wedge   {\cal F}_1 + 
     e^{-\varphi_2 -\varphi_3}\, {* {\cal F}_2}\wedge   {\cal F}_2\big)\nn\\
&- \chi_1\, (  F_{1}\wedge  {\cal F}_1 + 
                   F_{ 2}\wedge  {\cal F}_2)\,,
\label{d4lag}
\end{align}
where the index $i$ labelling the dilatons $\varphi_i$ and axions $\chi_i$
takes the values $i= 1,2,3$.  The four U(1)  field strengths can be written in 
terms of potentials as
\begin{eqnarray}
  F_{ 1} &=& d   A_{1} - \chi_2\, d {\cal A}_2\,,\nn\cr
  F_{ 2} &=& d  A_{ 2} + \chi_2\, d {\cal A}_1 - 
    \chi_3\, d   A_{ 1} +
      \chi_2\, \chi_3\, d  {\cal A}_2\,,\nn\cr
  {\cal F}_1 &=& d  {\cal A}_1 + \chi_3\, d  {\cal A}_2\,,\nn\cr
  {\cal F}_2 &=& d  {\cal A}_2\,.
\end{eqnarray}
 
 The general four-dimensional axisymmetric black hole metric is given by
\begin{equation}
d  s^2  = -  \Delta^{-1/2}  G \, 
( d{ t}+{ {\cal  A\, \mathrm{d}\phi}})^2 + { \Delta}^{1/2}
\left(\frac{d r^2} { X} + 
d\theta^2 + \frac{ X}{  G} \sin^2\theta\, d\phi^2 \right)\,.\label{metric4d}
\end{equation}
Here the quantities $X, G, {\cal A},  \Delta $ are all functions of $r$ and $\sin \theta$ only (and depend on  the mass, rotation and charge parameters). The first three are the same for the original and the corresponding subtracted black hole,  while the difference in $\Delta$ is the hallmark of subtracted geometry. (The function $\Delta(r,\theta)$ is called the warp factor of the black hole geometry.) The physical parameters (mass $M$, angular momentum $J$ and charges $Q_I$) of the general four-charge black hole are parametrized in terms of auxiliary constants $m, a, \delta_I$ as:
\begin{eqnarray}
G_4 M  &= &{\frac{1}{4}}m\sum_{I=0}^3\cosh 2\delta_I ~, \label{Mdef} \nonumber\\
G_4 Q_I & = &{\frac{1}{4}}m\sinh 2\delta_I~,~(I=0,1,2,3)~,\nonumber\\
G_4 J & = &m\, a \,(\Pi_c - \Pi_s)~,
\end{eqnarray}
where $G_4$ is the four-dimensional Newton constant and we employ the abbreviations
\begin{equation}
\Pi_c \equiv \prod_{I=0}^3\cosh\delta_I 
~,~~~ \Pi_s \equiv  \prod_{I=0}^3 \sinh\delta_I~.
\end{equation}
The functions  $X, G, {\cal A} $ are given by:
\begin{eqnarray}
{ X} & =& { r}^2 - 2{ m}{ r} + { a}^2~,\cr
{ G} & = &{ r}^2 - 2{ m}{ r} + { a}^2 \cos^2\theta \cr
&=& X-a^2\sin^2\theta\,,\cr
{ {\cal A}}  &=&{2{ m} { a} \sin^2\theta \over { G}}
\left[ ({ \Pi_c} - { \Pi_s}){  r} + 2{ m}{ \Pi}_s\right]\cr
 &=&{{a\sin^2\theta A_{red}}\over G}\,, \label{c4d}
\end{eqnarray}
where $A_{red}= 2m [(\Pi_c - \Pi_s)r + 2m \Pi_s]$. For the original black hole solutions, the remaining function $\Delta=\Delta_0$ is given by:
\begin{align}\label{delta0}
{ \Delta}_0 &= \prod_{I=0}^4 ({ r} + 2{ m}\sinh^2 { \delta}_I)
+ 2 { a}^2 \cos^2\theta \left[{ r}^2 + { m}{ r}\sum_{I=0}^3\sinh^2{ \delta_I}
+\,  4{ m}^2 ({ \Pi}_c - { \Pi}_s){ \Pi}_s \right. \nonumber\\ 
& \left.-  2{ m}^2 \sum_{I<J<K}
\sinh^2 { \delta}_I\sinh^2 { \delta}_J\sinh^2 { \delta}_K\right]
+ { a}^4 \cos^4\theta\,,
\end{align}

The scalars, axions and gauge fields for the most general case of all unequal charges was given in \cite{CCLPI}. Here we will only mention these quantities for the case of three equal charges and fourth unequal \cite{Cvetic:2012tr}, which was used in the scaling limit to obtain a subtracted geometry of the general four charge black hole. The scalar and axion fields are of the form:

\begin{eqnarray}
&&\chi_1=\chi_2=\chi_3= \frac{2{ m}\, { a}\cos\theta\, \cosh{ \delta} \sinh { \delta}(\cosh{ \delta}\sinh{ \delta}_0-\sinh { \delta}\cosh {\delta} _0)}{({ r}+2{ m}\sinh^2 { \delta})^2+ { a}^2\cos^2\theta},\cr
&&e^{\varphi_1} =e^{\varphi_2} =e^{\varphi_3} = \frac{({ r}+2{\tilde m}\sinh^2 { \delta})^2+ { a}^2\cos^2\theta}{{ {\Delta}_0}^{1\over 2}}\, ,
\end{eqnarray}

and the gauge fields are given by,

\begin{eqnarray}
A = && \frac{2 m}{ \Delta_0}\, \left\{
  \left[(r+2{ m}\sinh^2 { \delta})^2({ r}+2{ m}\sinh^2 { \delta}_0)+{ r}{ a}^2\cos^2\theta\right]\left[\cosh { \delta}\sinh { \delta} \, d{ t} \right.\right.\cr
 &&\left.  -{ a}\sin^2\theta
\cosh { \delta}\sinh { \delta}(\cosh { \delta}\cosh { \delta}_0-\sinh { \delta}\sinh { \delta}_0)\, d\phi\right] \cr
&&\left.+ 2{ m} \,{ a}^2\cos^2\theta \,\left[e \, d{ t} - { a}\sin^2\theta\sinh^2 { \delta}\cosh { \delta}\sinh { \delta}_0\,
  d\phi\right]\right\}\, , \cr
{\tilde{A}} = &&\frac{2 m}{ { \Delta_0}}\, \left\{
  \left[({ r}+2{ m}\sinh^2 { \delta})^3+{ r}{ a}^2
 \cos^2\theta\right]\left[\cosh{\delta}_0\sinh{ \delta}_0 \, d{ t}\right.\right.\cr 
 &&  \left.-{ a}\sin^2\theta(
\cosh^3{ \delta}\sinh{ \delta}_0 -\sinh^3{ \delta}\cosh { \delta}_0)\, d\phi\right] \cr
&&\left.+ 2{ m} \,{ a}^2\cos^2\theta \,\left[e_0 \, d{ t} -{ a} \sin^2\theta\sinh^3 { \delta}\cosh { \delta}_0\,
  d\phi\right]\right\}\,.
  \end{eqnarray}
Here:
\begin{eqnarray}
e&=&\sinh^2 { \delta}\cosh^2 { \delta}\cosh { \delta}_0\sinh { \delta}_0(\cosh^2 { \delta}+\sinh^2 { \delta})\cr &&-\sinh^3 { \delta}\cosh { \delta} (\sinh^2 { \delta}+2\sinh^2 { \delta}_0+2\sinh^2 { \delta}\sinh^2 { \delta}_0),\cr
e_0&=&\sinh^3 { \delta}\cosh^3 { \delta}(\cosh^2 { \delta}_0+\sinh^2 { \delta}_0)-\sinh { \delta}_0\cosh { \delta}_0(3\sinh^4 { \delta}+2\sinh^6 { \delta})\,,
\end{eqnarray}
and $A_1=A_2=A_3\equiv A $ are the gauge fields for gauge field strengths    $* F_1= F_2= * {\cal F} _1\equiv F\, $,  and  $A_4\equiv {\tilde{A}}$ is the gauge field for  $ {\cal F}_2\equiv {\cal F}$. 

For the case of subtracted geometry we replace the function $\Delta_0$ by $\Delta_{sub}$ \cite{Cvetic:1997uw}, given by:

\begin{equation}\label{deltasub}
\Delta_{sub} = (2 m)^3 r ( \Pi_c^2 -\Pi_s^2) + (2m)^4 \Pi_s^2 - (2m)^2 ( \Pi_c-\Pi_s)^2 a^2 \cos^2 \theta\,.
\end{equation}

The matter fields for the subtracted geometry black holes were obtained in \cite{Cvetic:2012tr} using the scaling limit and are,
\begin{eqnarray}
&&\chi_1=\chi_2=\chi_3=-\frac{a(\Pi_c-\Pi_s)\cos\theta}{2m} , \ \ e^{\varphi_1} =e^{\varphi_2} =e^{\varphi_3} = \frac{(2m)^2} {\Delta_{sub}^{1\over 2}}\, ,\label{canonicalsc}
\end{eqnarray}
and
\begin{eqnarray}
A=&&-\frac{r}{2m}dt+\frac{(2m) a^2[2m\Pi_s^2 -r (\Pi_c-\Pi_s)^2]\cos^ 2\theta}{\Delta_{sub}}dt\, \cr\
&&-{ a(\Pi_c-\Pi_s)\sin^2\theta}(1+\frac{(2m)^2a^2(\Pi_c-\Pi_s)^2\cos^2\theta}{\Delta_{sub}})\, d\phi ,
\cr
{\cal A}= &&\frac{ (2m)^4 \Pi_c\Pi_s + (2m)^2a^2 (\Pi_c-\Pi_s)^2\cos^2\theta}{(\Pi_c^2-\Pi_s^2)\Delta_{sub}}\, dt \, + \frac{(2m)^4a(\Pi_c-\Pi_s)\sin^2\theta}{\Delta_{sub}}   \, d\phi\, .\label{gaugeps}
\end{eqnarray}

In both the original and the subtracted case the horizons, specified by $X=0$, are at:
\begin{equation}
r_\pm=m\pm \sqrt{m^2-a^2}\, .
\end{equation}
The outer horizon  $\Sigma$ is the thus the two-dimensional surface at fixed $r,t$ defined by $r=r_+, t=\mathrm{const}$. The general formula for its area is
\begin{equation}\label{horarea}
A_{\Sigma}= \int d\theta d\phi\, \sin \theta \,A_{red}\big|_{r=r_+}\,.
\end{equation}
Clearly the area is unmodified by the change $\Delta_0\rightarrow\Delta_{sub}$, and thus the Bekenstein-Hawking entropy of the subtracted black hole equals that of the original one:
\begin{equation}
S_{BH}^{(sub)}= \frac{A_\Sigma^{(sub)}}{4G}= \frac{A_\Sigma^{(or)}}{4G}=S_{BH}^{(or)}\,.
\end{equation}

\section{Entanglement entropy of original and subtracted black holes}

\subsection{Black hole entanglement entropy}

Entanglement entropy of quantum fields, computed across the black hole event horizon $\Sigma$, gives a divergent expression of the form
\begin{equation}
S^{ent}\sim\frac{A_\Sigma}{\epsilon^2}+c_0 \ln\left(\frac{L}{\epsilon} \right)+S_{finite}\,,
\end{equation}
where $\epsilon$ is a short-distance UV cutoff and $L$ an IR cutoff. It is well known \cite{Solodukhin:2011gn} that the divergences in this expression match the divergences in the one-loop effective action for quantum fields in the black hole background. This means that when we view the total black hole entropy as composed of a ``bare gravitational'' or ``tree-level''  entropy $S^{tree}$, plus the entanglement entropy as a ``quantum correction'' $S^{loop}$, then the total entropy $S_{BH}^{(tot)}$ takes the same general form as $S^{tree}$ with the one-loop renormalized couplings replacing the bare couplings present in $S^{tree}$. These couplings which renormalize are the Newton constant $G_4$, and couplings $c_{1,2,3}$ for higher-order curvature terms $R^2$, $R_{\mu\nu}R^{\mu\nu}$, and $R_{\lambda\mu\nu\rho}R^{\lambda\mu\nu\rho}$ added to the Lagrangian. 
(The assumption of a ``bare'' gravitational entropy can be disposed of in a specific model in which gravity is effective and wholly induced by quantum fields \cite{Fursaev:1995ef},\cite{Sakharov:1967pk}.)

Both the tree-level entropy and the loop corrections  are computed with the conical singularity method. A  Euclidean
manifold is obtained by Wick rotation of the Lorentzian black hole geometry. 
One creates a conical defect around the horizon (giving  periodicity $2\pi\alpha$ to the Euclidean time coordinate, which loops around it). The tree-level entropy is then obtained from the expression:
\begin{equation}
S^{tree} = \left(\alpha\partial_\alpha -1\right) S_\alpha^B|_{\alpha=1}\,,
\end{equation}
where $S_\alpha^B$ is the bare gravitational action, including higher-order curvature terms, evaluated on the  conical Euclidean manifold. The loop correction is given by the same equation but replacing the bare gravitational action by minus the log of the quantum partition function:
\begin{equation}
S^{loop} = -\left(\alpha\partial_\alpha -1\right) \ln Z_\alpha\Big|_{\alpha=1}
\end{equation}
These expressions have been computed for Kerr-Newman black holes by Solodukhin and Mann \cite{SOM} and for arbitrary axisymmetric black holes by Jing and Yan \cite{jing}.
They are respectively given by
\begin{equation}
S^{tree}=\frac{A_{\Sigma}}{4G_4^B}-8\pi \int_{\Sigma}
 \left[\left(c^1_{B}R+\frac{c^2_{B}}{2}\sum^2_{a=1}R_{\mu
\nu}n^\mu _in^\nu _i +
c^3_{B}\sum^2_{a,b=1}R_{\mu \nu \alpha \beta }n^\mu _in^\nu _j
n^\alpha _i n^\beta _j\right)\right], \label{tree}
\end{equation}

\begin{align} S^{loop}&=\frac{A_{\Sigma}}{48\pi \epsilon
^2}+\Bigg\{\frac{1}{144\pi} \int_{\Sigma}
R-\frac{1}{45}\frac{1}{16\pi}\int_{\Sigma} \Big(
\sum^2_{a=1}R_{\mu \nu}n^\mu _in^\nu _i-2\sum^2_{a,b=1}R_{\mu \nu
\alpha \beta }n^\mu _in^\nu _j n^\alpha _i n^\beta _j\Big)
\nonumber\\
\\ & -\frac{1}{90}\frac{1}{16\pi}\int_{\Sigma}
\left(K^aK^a\right)+
\frac{1}{24\pi}\Big(\lambda_1-\frac{\lambda_2}{30}\Big)
\int_{\Sigma}\left(K^aK^a-2tr(K.K)\right) \Bigg\}
\ln\frac{L}{\epsilon}, \label{solo0}
\end{align}
where  $G_4^B, c^I_{B}, (I=1,2,3)$ represent bare constants
(tree-level), $K^a_{\mu\nu}=-\gamma^{\alpha}_{\mu}
\gamma^{\beta}_{\nu}\nabla_{\alpha}n^a_{\beta}$ is the extrinsic
curvature, $K^a=g^{\mu \nu }K^a_{\mu \nu}$ is the trace of the
extrinsic curvature, and $n_i^\mu$ ($i=1,2$) are unit vectors normal to $\Sigma$. The notation $\int_\Sigma$ stands for the right hand side of (\ref{horarea}).

For the general axisymmetric black holes all the quantities dependent on the extrinsic curvature vanish, thus verifying that the tree-level and the loop formulas have the same general form and the entropy renormalizes. We quote from \cite{jing} a useful expression for the  combination of Riemann tensor contractions: 
\begin{align}
R_{nn}(r_+,\theta)-2R_{mnmn}(r_+,\theta)=&
\Bigg\{\frac{\partial ^2g^{rr}}{\partial r^2}+
\frac{3}{2}\frac{\partial g^{rr}}{\partial r}\frac{\partial \ln
f}{\partial r}-\frac{1}{2}\frac{\partial g^{rr}}{\partial
r}\left(\frac{1} {g_{\theta \theta}}\frac{\partial g_{\theta
\theta}}{\partial r} +\frac{1}{g_{\varphi \varphi}}\frac{\partial
g_{\varphi \varphi}}{\partial r}\right) \nonumber \\ & 
-\frac{2g_{\varphi
\varphi}}{f} \left[\frac{\partial}{\partial r}\left(\frac{g_{t
\varphi}}{g_{\varphi
\varphi}}\right)\right]^2\Bigg\}_{r_+}\,.\nonumber \\ \label{a5}
\end{align}
Here  the
Boyer-Lindquist form of the Euclidean metric 
\begin{equation}
ds^2=g_{tt}dt^2+g_{rr}dr^2+ 2 g_{t \varphi}dtd\varphi+g_{\theta
\theta}d\theta ^2+g_{\varphi \varphi}d\varphi ^2,
 \label{metric0} \end{equation}
 is assumed, with
 $g_{tt}$, $g_{rr}$, $g_{t\varphi}$, $g_{\theta \theta}$ and
$g_{\varphi \varphi}$ functions of the coordinates $r$ and
$\theta$. The inverse metric component is $ g^{rr}=1/g_{rr}$, and $ f= - g_{rr}\left(g_{tt}-\frac{g_{t\varphi}^2}{g_{\varphi\varphi}}
\right)$.

Note that because $R$ vanishes on the black hole metrics (\ref{metric4d}), the higher-order correction to the entropy is given essentially by the combination (\ref{a5}) integrated over the horizon. Replacing the metric components of (\ref{a5}) and evaluating on the horizon we obtain:
\begin{align}\label{Sloop}
S^{loop} &=\frac{A_{\Sigma}}{48\pi \epsilon
^2}- \frac{1}{720 \pi}\int_\Sigma \, \left(R_{nn}(r_H,\theta)-2R_{mnmn}(r_H,\theta) \right)\nonumber \\
&=\frac{A_{\Sigma}}{48\pi \epsilon
^2} -\frac{1}{720 \pi}\int_\Sigma \left( \frac{X''}{\Delta^{1/2}} + \frac{X' \,\Delta'}{ 2 \Delta^{3/2}} -  2 \frac{G' X'}{ G \Delta^{1/2}} - 2 \frac{ G^2{( \mathcal{A}')}^{2}}{ \Delta^{{3}/{2}} \sin^2 \theta} 
\right)_{r_+}\,,
\end{align}
where a prime stands for a derivative with respect to $r$.

 In the next subsections we give the results form this expression both for the original and the subtracted black hole geometries.

\subsection{Results for original black holes}
The evaluation of (\ref{Sloop})  for the original black hole geometry that has $\Delta=\Delta_0$ as given in (\ref{delta0}), in the fully general case with four charges and rotation that is parametrized by $(m,a,\delta_I)$, is given by an expression of the form
\begin{equation}\label{generaloriginal}
S^{loop} = \frac{A_{\Sigma}}{48 \pi \epsilon^2}-\frac{A_{red}(r_+)}{720}\int_{-1}^{1}\mathrm{d}u\, \frac{\kappa u^4+\lambda u^2+\mu}{(\alpha u^4 + \beta u+\gamma)^{3/2}}\,,
\end{equation}
where the six parameters $(\alpha,\beta, \gamma, \kappa, \lambda, \mu)$ depend on the black hole parameters $(m,a,\delta_I)$ (as do, of course, the horizon radius $r_+$ and the function $A_{red}(r_+)$ defined above in (\ref{c4d})). The definitions of these six parameters are given in the Appendix. The expression is obtained through the change of variables $u=\cos\theta$. The integral it features is in general expressible as a lengthy combination of elliptic functions, which can take different forms in different regions of parameter space. For this reason we will provide here only the results in some particular cases of physical interest.

The expression derived in \cite{SOM} for the entanglement entropy of the Kerr-Newman black hole (with rotation and a single charge parameter) is obtained in the limit $\delta_0=\delta_1=\delta_2=\delta_3\equiv\delta$. It reads:
\begin{equation}\label{originalKN}
S_{KN}^{loop} = \frac{A_{\Sigma}}{48 \pi \epsilon^2}+\frac{1}{45}\left[ 
1-\frac{3m^2\sinh^2(2\delta)}{4R_{+}^2} \left( 
1+\frac{(a^2+R_{+}^2)  \arctan \frac{a}{R_{+}} } {a\,R_{+}} 
\right)
  \right]  \log( \frac{r_+}{\epsilon}) \,,
\end{equation}
where
\begin{equation}
R_{+}= r_+ + 2 m \sinh ^2 \delta\,.
\end{equation}

The correspondence between this expression and  the result given by formula (4.12) in \cite{SOM} is manifest if we translate suitably our notation to the one used in this reference. For ease of comparison we provide the following translation for the notations, where the left hand side correspond to the notations used in \cite{SOM} and the right hand side to those used in the present work :

\begin{eqnarray}
q  &\longleftrightarrow & m \sinh(2\delta)\,,\\
m & \longleftrightarrow & m \cosh(2\delta)\,,\\
r_+ & \longleftrightarrow & r_+ + 2m \sinh^2\delta =R_{r_+}\,.
\end{eqnarray}

The results for the Reissner-Nordstrom and Schwarzschild black holes are obtained from the previous formula setting $a=0$ and $a=0=\delta$ respectively. They read:

\begin{eqnarray} 
S_{RN}^{loop} &=& \frac{A_{\Sigma}}{48 \pi \epsilon^2}+\frac{ 1}{90}  \frac{ ( 2 - \sinh^2 \delta)}{\cosh^2 \delta} \log\left( \frac{r_+}{\epsilon}\right)\,,\label{originalRN} \\
S_{Sch}^{loop} &=& \frac{A_{\Sigma}}{48 \pi \epsilon^2}+\frac{ 1}{45} \log\left( \frac{r_+}{\epsilon}\right)\,.\label{originalSch} 
\end{eqnarray}

On the other hand, the result for the static black hole with four different charges $(a=0, \delta_{0,1,2,3}\neq 0)$ reads

\begin{align} \label{original4q} 
S_{4q}^{loop} &= \frac{A_{\Sigma}}{48 \pi \epsilon^2}\\ \nonumber & +\frac{1}{360\,  \Pi_c}  \left(8( \Pi_c^2 -\Pi_s^2) - 3 \sum  s_I^2 -6 \sum_{I\neq J} s_I^2 s_J^2  - 9  \sum_{I<J<K}
s_I^2  s_J^2 s_K^2  -4 \prod_I s_I^2 \right)\log\left( \frac{r_+}{\epsilon}\right)\,, 
\end{align}
where $s_I = \sinh \delta_I$ and $I=0,1,2,3$. This result does not feature previously in the literature. It reduces to (\ref{originalRN}) when we set $\delta_I=\delta$ for all $I$. Note that each of the results (\ref{originalRN}, \ref{originalSch}, \ref{original4q}) has a log correction independent of the parameter $m$.

\subsection{Results for subtracted black holes}

We turn now to the evaluation of the entropy for the black holes with subtracted geometry. As remarked before, the black hole horizon area $A_\Sigma$ is independent of $\Delta$ and therefore the leading order term of the entropy always matches the original one. We will show that this agreement is not preserved for the subleading order, i.e. the logarithmic correction involving the integral of (\ref{a5}). 

The black hole entanglement entropy for subtracted geometry is computed by evaluating (\ref{Sloop}) with $\Delta=\Delta_{sub}$ as given by (\ref{deltasub}). The result for the fully general four-charge black hole with rotation is given by:

\begin{eqnarray} \label{generalsub}
&&S^{loop}=\frac{A_\Sigma}{48\pi\epsilon^2} \\
&&-\frac{1}{180}\frac{m^3(\Pi_c+\Pi_s)^3+\sqrt{(m^2-a^2)^{3}}(\Pi_c-\Pi_s)^3}{m[m\, ( \Pi_c + \Pi_s) + \sqrt{ m^2 -a^2 } ( \Pi_c - \Pi_s)] [m( \Pi_c^2 + \Pi_s^2) + \sqrt{ m^2 -a^2}  ( \Pi_c^2 - \Pi_s^2)]}\log\left(\frac{r_+}{\epsilon}\right)\,. \nonumber
\end{eqnarray}

The result at the subleading order is clearly different from the original black hole expression (\ref{generaloriginal}), which is much more complex and depends on all four charge parameters $\delta_I$ separately instead of through the combinations $\Pi_c,\Pi_s$. For completeness we include below the results for the subtracted versions of the Kerr-Newmann black hole, the Reissner-Nordstrom black hole, the Schwarzschild  and the static four-charge black hole. These results are to be contrasted with the original expressions (\ref{originalKN}, \ref{originalRN}, \ref{originalSch}, \ref{original4q}).
\begin{eqnarray}
S_{KN-sub}^{loop} &=&  \frac{A_{\Sigma}}{48 \pi \epsilon^2}\\ &-&\frac{ 1}{180} \frac{ m^3 ( \cosh ^4 \delta  + \sinh ^4 \delta)^3 + \sqrt{ ( m^2 -a^2)^3} ( \cosh ^4 \delta  - \sinh ^4 \delta)^3}{ m ( \tilde \gamma_1 \tilde \gamma_2)} \,,   \label{subKN} \\
S_{RN-sub}^{loop} &=& \frac{A_{\Sigma}}{48 \pi \epsilon^2}-\frac{ 1}{360}  \frac{(\cosh^8 \delta + 3 \sinh^8 \delta)}{ \cosh^8 \delta}   \log\left( \frac{r_+}{\epsilon}\right) \,, \label{subRN}  \\
S_{Sch-sub}^{loop} & = & \frac{A_{\Sigma}}{48 \pi \epsilon^2}-\frac{ 1}{360} \log\left( \frac{r_+}{\epsilon}\right) \,, \label{subSch} \\
S_{4q-sub}^{loop} &=& \frac{A_{\Sigma}}{48 \pi \epsilon^2} -\frac{ 1}{360}  \left(\frac{(\Pi_c^2 + 3 \Pi_s^2}{\Pi_c^2}\right) \log\left( \frac{r_+}{\epsilon}\right) \,. \label{sub4q} 
\end{eqnarray}

In (\ref{subKN}), $\tilde{\gamma_1 }$ stands for $m\, ( \cosh^4\delta + \sinh^4\delta) + ( m^2 -a^2 )^{1/2} ( \cosh^4\delta-\sinh^4\delta)$ and $\tilde{\gamma_2 }$ stands for $m\, ( \cosh^8 \delta + \sinh^8\delta) + ( m^2 -a^2 )^{1/2} ( \cosh^8\delta-\sinh^8\delta)$. These expressions are all obtained evaluating (\ref{generalsub}) in the appropriate limits.  Just as before, the static results (\ref{subRN}-\ref{sub4q}) have the log prefactor independent of $m$. It is seen, however, that they in every case the expression is different from the corresponding expression for the original black hole.

Nevertheless, there is a limit in which the expressions coincide. The subtracted geometry is designed to be a modification of the original black hole geometry that preserves the key features of extremal black holes even for non-extremal parameters. Therefore in the extremal limit the entropies of the original and the subtracted black holes should match exactly. In our parametrization, this limit is given by:
\begin{equation}
m\longrightarrow 0\,,\quad \delta_I \longrightarrow +\infty\,,\quad\quad \mathrm{with}\,\,\, m\exp (2\delta_I) = 4G_4\,Q_I = \mathrm{finite}\,.
\end{equation}

When taking this limit, we find indeed agreement between the original and the subtracted entropies for extremal black holes:
\begin{equation}
S_{ext}^{loop} =S_{ext-sub}^{loop} =  \frac{A_{\Sigma}}{48 \pi \epsilon^2}-\frac{ 1}{90}   \log\left( \frac{r_+}{\epsilon}\right)\,.
\end{equation}

\section{Interpolating Schwarzschild geometry and vanishing log correction}

If we compare the results (\ref{originalSch}) and (\ref{subSch}), which express the entanglement entropy for the original and subtracted versions of the Schwarzschild black hole respectively, we notice an interesting feature: the sign of the logarithmic correction has changed from positive to negative. This raises the question of whether there exists an interpolating geometry for which this correction vanishes. One could interpret such a solution as a fixed point for the entropy, in the sense of the renormalization group, since the tree-level result is unaffected by the log correction. 

As it happens, we can indeed construct solutions that interpolate between the original black hole geometry and the subtracted geometry. This is done through Harrison transforms, which are a four-parameter group of transformations acting on the black hole solution. It is shown in \cite{Cvetic:2013cja} that a four-parameter Harrison transform interpolates between the original geometry and a new black hole geometry, which after a rescaling yields the subtracted version of the original geometry. (A version of this construction requiring only two Harrison parameters had appeared previously in \cite{Cvetic:2012tr}.) Therefore, one should expect to find a suitable combination of Harrison transformation parameters that corresponds to a geometry with a vanishing log term in the entropy.

Let us review briefly how the general Harrison transform is defined in \cite{Cvetic:2013cja}. We have four Harrison parameters $(\alpha_0, \alpha_j)$, with $j=1,2,3$. The effect of a Harrison transformation $g_H(\alpha_0,\alpha_j)$ on the black hole geometry is to modify the warp factor $\Delta$ (which is, in general, a polynomial of the fourth order in $r$) in the following way: The term with $r^4$ gets multiplied by $(1-\alpha_0^2)\prod_j(1-\alpha_j^2)$, so that it vanishes when any of the four parameters is set to 1. The term with $r^3$ is multiplied by a suitable permutation of terms combining three factors of the form $(1-\alpha^2)$, so that it vanishes when any two of the parameters are set to 1. The analogous operation happens with the second and first order terms. Detailed formulas, too long to quote here, are to be found in Appendix C of \cite{Cvetic:2013cja} for the particular cases of the Kerr solution and the general static solution.

To obtain the subtracted geometry from the original geometry, one needs to apply a Harrison transform with particular values of the four parameters, followed by a specific re-scaling of the metric. The values of the Harrison parameters that lead to the subtracted geometry are:
\begin{equation}
\alpha_j = 1\,,\quad\quad\alpha_0=\frac{\Pi_s\cosh\delta_0-\Pi_c\sinh\delta_0}{\Pi_c\cosh\delta_0-\Pi_s\sinh\delta_0}\,,
\end{equation}
and the subsequent re-scaling of the metric takes the form:
\begin{equation}
U\to U+c_0\,,\quad\quad \mathrm{e}^{2 c_0}=\frac{\mathrm{e}^{\delta_1+\delta_2+\delta_3}}{\Pi_c\cosh\delta_0-\Pi_s\sinh\delta_0}\,,
\end{equation}
where $\exp(-4U)=\Delta/G^2$. The matter fields supporting the geometry get rescaled as well; we omit these details for briefness and refer the reader once more to \cite{Cvetic:2013cja} for the full formulas.

To move between the original and the subtracted versions of Schwarzschild, there is no metric re-scaling involved because all $\delta_I$ vanish and therefore so does $c_0$. Also, in this case both the initial and final values of the $\alpha_0$ parameter are 0, so we may disregard it. The interpolating geometry we have is given by 
\begin{equation}
d  s^2  = -  \Delta^{-1/2}  G \, 
dt^2 + { \Delta}^{1/2}
\left(\frac{d r^2} { X} + 
d\theta^2 + \frac{ X}{  G} \sin^2\theta\, d\phi^2 \right)\,,
\end{equation}

with 
\begin{equation}
\Delta= \prod_I (( 1- \alpha_I^2)r  + 2m \alpha_I^2).
\end{equation}

We can now compute the entropy directly for the interpolating geometries using (\ref{Sloop}). The lack of angular dependence makes the calculation trivial, and the result is

\begin{equation}
S^{loop}_{interpolating} = \left( \frac{ 3 \alpha_0^2 + 3 \alpha_1^2 + 3 \alpha_2^2 + 3 \alpha_3^2 -8}{ 45}\right) \log \left( \frac{r_+}{\epsilon}\right).
\end{equation}

One can easily see that 

\begin{eqnarray}
S^{loop}_{interpolating} &=& S_{Sch}^{loop}\;\; \;\;\;\;\; \; \;\; \textrm{for $\alpha_{0,1,2,3} =0$} ,\\ \nonumber
 &=& S_{Sch-sub}^{loop} \;\; \;\; \textrm{for $\alpha_{1,2,3}=1$ , $\alpha_0=0$} ,\\ \nonumber
  &=& S_{ext}^{loop} \;\;\;\;\;\;\; \; \;\; \textrm{for $\alpha_{0,1,2,3}=1$} ,\\ \nonumber
   &=& 0\;\;\;\;\;\;\; \;\;\;\;\; \; \;\; \textrm{for $\sum_I \alpha_I^2=\frac{8}{3}$ }.\\ \nonumber
  \end{eqnarray}
  We conclude that a combination of Harrison parameters satisfying  $\sum_I \alpha_I^2=\frac{8}{3}$ takes us from the original Schwarzschild black hole to one with vanishing logarithmic corrections to the entropy.
  
  \section{Conclusion and Future Directions}
  
  We have studied the logarithmic corrections to the entanglement entropy of a minimally coupled scalar field in the subtracted geometry black hole background. Our main results are collected in formulas (\ref{generalsub}-\ref{sub4q}). They all differ from the corresponding results for non-subtracted black holes, indicating that the agreement of subtracted and non-subtracted entropies does not extend beyond the tree level. On the other hand, the subtracted results approach the original ones for the extremal BPS case in the appropriate limit. We noticed that the logarithmic correction term universally changes sign for all cases of subtracted black holes. For the schwarzchild case we found the interpolating solution which for certain choices of the Harrison parameters gives a vanishing logarithmic correction.
  
 An interesting direction would be to understand the contributions of other spin fields to the logarithmic corrections of the entanglement entropy of the subtracted geometry black holes. In the last few years a lot of work has been done on calculating and understanding the logarithmic corrections to black hole entropy using the euclidean gravity (or heat kernel) method. For a review of the various different backgrounds studied see \cite{Solodukhin:2011gn},\cite{Sen:2012dw}. It will be interesting to study the subtracted geometry black holes with different spin contributions using this machinery. \\
  
 We have also used the Harrison transformations to study the entanglement entropy of the interpolating geometry that connects the original schwarzchild black hole to the subtracted one. We have found the choice of the Harrison parameters for which the logarithmic corrections vanish. Since we know that the Harrison parameters correspond to turning on irrelevant operators in the CFT interpretation when the geometry is lifted to one higher dimension, it would be interesting to find out the CFT interpretation of the particular point in the Harrison orbit for which the logarithmic corrections vanish. One may see that this particular point has a relationship with a fixed point in the renormalization group flow of the conformal field theory where the irrelevant operators are turned on.\\
 
 Also, recently the most general black hole solution of the STU model (a full generating solution for N=4 and N=8 supergravity theory) has been found \cite{chow},\cite{chow1}. The subtracted geometry of these black holes has been studied in \cite{mirjamlarsen}. It will be enlightening to study the entanglement entropy and logarithmic corrections in this setting. Finally, it will also be interesting to study the entanglement entropy of the subtracted geometry black holes in one dimension higher where the geometry is $AdS_3\times S^2$ and the symmetries of the maximally symmetric spaces could be exploited using the heat kernel approach.

  \acknowledgments
  We would like to thank G. Gibbons, M. Guica and F. Larsen for useful discussions and comments. The work is supported in part by the DOE Grant DOE-EY-76-02-
3071, the Fay R. and Eugene L. Langberg Endowed Chair, the Slovenian 
Research Agency (ARRS), and the Simons Foundation Fellowship.
  
  \appendix
\section{Parameters for the  original general black hole expression}
  
  In this Appendix we provide the definitions of the parameters in the general expression for the entropy of original four-charge rotating black holes. We quote again formula (\ref{generaloriginal}):
 \begin{equation}
S^{loop} = \frac{A_{\Sigma}}{48 \pi \epsilon^2}-\frac{A_{red}(r_+)}{720}\int_{-1}^{1}\mathrm{d}u\, \frac{\kappa u^4+\lambda u^2+\mu}{(\alpha u^4 + \beta u+\gamma)^{3/2}}\,,
\end{equation} 

The denominator is simply $\Delta_0^{3/2}$ as given in (\ref{delta0}) with the change of variables $u=\cos\theta$. Hence we have:

\begin{eqnarray}
\alpha & = & a^4,\\
\beta & = & 2  a^2 (r_+^2 +C),\\
\gamma & = &  \prod_{I=0}^4(R_+^I)^2,
\end{eqnarray}
  where $R_+^I= r_+ + 2m \sinh^2 \delta_I$ and
  \begin{equation}
 C=  { m}r_+\sum_{I=0}^3\sinh^2{ \delta_I}
+\,  4{ m}^2 ({ \Pi}_c - { \Pi}_s){ \Pi}_s -  2{ m}^2 \sum_{I<J<K}
\sinh^2 { \delta}_I\sinh^2 { \delta}_J\sinh^2 { \delta}_K\,.
\end{equation}

The remaining three parameters are given by:

\begin{eqnarray}
\kappa & = & 4a^4\,,\\
\lambda & = &4\beta-16 a^2 m^2\left(\Pi_c-\Pi_s\right)^2 +2a^2(r_+-r_-)(2r_+-2r_--R_b)\,,
\end{eqnarray}
and
\begin{align}
\mu  =&\,  4\gamma - 16 a^2 m^2\left(\Pi_c-\Pi_s\right)^2\nonumber\\
&+\left[16m A_{red}(r_+)(\Pi_c-\Pi_s)+R_a+4(r_+^2+2m\,r_++2C) \right](r_+-r_-)\,,
\end{align}
where we use the abbreviations
\begin{equation}
R_a = R^0_+ R^1_+ R^2_+ + R^0_+R^1_+R^3_+ + R^0_+R^2_+R^3_+ + R^1_+R^2_+R^3_+ ,
\end{equation}
 and 
\begin{equation}
R_b= 2 r_+ +  m \sum_I \sinh^2 \delta_I .
\end{equation}

\end{document}